\documentclass[a4paper,fleqn,usenatbib]{mnras}
\usepackage{newtxtext,newtxmath}
\usepackage[T1]{fontenc}
\usepackage{ae,aecompl}

\usepackage{graphicx}
\usepackage{amsmath}
\usepackage{amssymb}
\usepackage{footnote}

\def\wid{WASP-174}
\def\widb{WASP-174b}
\def\prot{$P_{\rm rot}$}
\def\porb{$P_{\rm orb}$}
\def\teff{$T_{\rm eff}$}

\title[WASP-174b]{Discovery of WASP-174b: Doppler tomography of a near-grazing transit}

\author[L. Temple et al.]{
L.Y. Temple,$^{1}$\thanks{E-mail: l.y.temple@keele.ac.uk}
C. Hellier$^{1}$,
Y. Almleaky$^{2,3}$,
D.R. Anderson$^{1}$,
F. Bouchy$^{4}$,\newauthor
D.J.A. Brown$^{5,6}$,
A. Burdanov$^{8}$,
A. Collier Cameron$^{7}$,
L. Delrez$^{9}$,
M. Gillon$^{8}$,
R. Hall$^{9}$,\newauthor
E. Jehin$^{8}$,
M. Lendl$^{10,4}$,
P.F.L. Maxted$^{1}$,
L. D. Nielsen$^{4}$,
F. Pepe$^{4}$,
D. Pollacco$^{5,6}$,\newauthor
D. Queloz$^{9}$,
D. S\'egransan$^{4}$,
B. Smalley$^{1}$,
S. Sohy$^{8}$,
S. Thompson$^{9}$,
A.H.M.J. Triaud$^{11}$,\newauthor
O.D. Turner$^{4,1}$,
S. Udry$^{4}$,
R.G. West$^{5}$
\\
$^{1}$Astrophysics Group, Keele University, Staffordshire, ST5 5BG, UK\\
$^{2}$Space and Astronomy Department, Faculty of Science, King Abdulaziz University, 21589 Jeddah, Saudi Arabia\\
$^{3}$King Abdullah Centre for Crescent Observations and Astronomy (KACCOA), Makkah Clock, Saudia Arabia\\
$^{4}$Observatoire astronomique de l'Universit\'e de Gen\`eve 51 ch. des Maillettes, 1290 Sauverny, Switzerland\\
$^{5}$Department of Physics, University of Warwick, Gibbet Hill Road, Coventry, CV4 7AL, UK\\
$^{6}$Centre for Exoplanets and Habitability, University of Warwick, Gibbet Hill Road, Coventry CV4 7AL, UK\\
$^{7}$SUPA, School of Physics and Astronomy, University of St.\ Andrews, North Haugh,  Fife, KY16 9SS, UK\\
$^{8}$Space sciences, Technologies and Astrophysics Research (STAR) Institute, Universit{\'e} de Li{\`e}ge, All{\'e}e du 6 Ao{\^u}t 17, 4000 Li{\`e}ge, Belgium\\
$^{9}$Cavendish Laboratory, J J Thomson Avenue, Cambridge, CB3 0HE, UK\\
$^{10}$Space Research Institute, Austrian Academy of Sciences, Schmiedlstr. 6, 80 42, Graz, Austria\\
$^{11}$School of Physics \& Astronomy, University of Birmingham, Edgbaston, Birmingham, B15 2TT, UK\\
}

\date{Accepted XXX. Received YYY; in original form ZZZ}

\pubyear{2018}

\begin{document}
\label{firstpage}
\pagerange{\pageref{firstpage}--\pageref{lastpage}}
\maketitle

\begin{abstract}
We report the discovery and tomographic detection of \widb, a planet with a near-grazing transit on a 4.23-d orbit around a $V$\,=\,11.9, F6V star with [Fe/H]\,=\,0.09\,$\pm$\,0.09.  The planet is in a moderately misaligned orbit with a sky-projected spin--orbit angle of $\lambda$\,=\,31$^{\circ}$\,$\pm$\,1$^{\circ}$. This is in agreement with the known tendency for orbits around hotter stars to be misaligned. Owing to the grazing transit the planet's radius is uncertain, with a possible range of 0.8--1.8\,R$_{\rm Jup}$. The planet's mass has an upper limit of 1.3\,M$_{\rm Jup}$.    \wid\ is the faintest hot-Jupiter system so far confirmed by tomographic means. 
\end{abstract}


\section{Introduction}
\label{sec:intro}
Hot-Jupiter exoplanets orbiting stars of A--mid-F spectral types, which lie beyond the Kraft break at \teff\,>\,6250\,K \citep{1967ApJ...150..551K}, are likely to have different properties from those orbiting cooler stars. First, planets with hot stars will be more highly irradiated, producing hotter and sometimes ``ultra-hot'' Jupiters (with T$_{\rm eq}\,\gtrsim$\,2200\,K). The high irradiation is also thought to be related to the inflated radii seen in many hot Jupiters \citep[e.g. ][]{2016AJ....152..182H}. Second, hot Jupiters transiting hotter stars are more likely to be in misaligned orbits, a trait which was first noted by \citet{2010ApJ...718L.145W} and has been discussed at length in recent literature \citep[e.g.][for a review, see \citet{2017haex.bookE...2T}]{2014ApJ...786..102V,2015ApJ...801....3M,2017AJ....153..205D}. And third, where early-type stars are fast rotators, the rotation period can be shorter than the planet's orbital period, giving a systematically different tidal interaction than in most hot-Jupiter systems \citep[see, e.g.,][]{2017AJ....153...94C}. It is important to understand these differences in order to create a complete picture of how planetary systems form and evolve.

Another difference between hot Jupiters and ultra-hot Jupiters is the efficiency with which heat can be transported from the tidally-locked dayside of the planet to its nightside. More highly-irradiated planets are less efficient at recirculating heat within the atmosphere: ultra-hot Jupiters have dayside temperatures close to the local radiative equilibrium temperature \citep{2015AREPS..43..509H}, indicative of inefficient transport of heat to the nightside.

The inflated radii of hot Jupiters makes them ideal candidates for studying planetary atmospheres using transmission spectroscopy \citep[e.g ][]{2015ApJ...814...66K,2015A&A...577A..62W,2017MNRAS.467.4591G,2018NatAs.tmp...86Y} while the high surface temperatures make it possible to observe the thermal emission of these planets in the infrared \citep[e.g.][]{2010A&A...511A...3G,2014ApJ...791...36S}.

High irradiation combined with absorption by molecules such as TiO and VO is expected to produce a thermal inversion in the upper atmosphere \citep{2008ApJ...678.1419F}, and this has been found in some ultra-hot Jupiters \citep[e.g. WASP-121b, ][]{2017Natur.548...58E} but not in others \citep[e.g. Kepler-13Ab, ][]{2017AJ....154..158B}.  In WASP-103b, \citet{2018AJ....156...17K} find an inversion on the irradiated day side of the planet but not on the cooler night side.   Water features can be prominent in the spectra of cooler hot Jupiters \citep[e.g. WASP-107b, ][]{2018ApJ...858L...6K}, but may be absent on the dayside of hotter planets such as WASP-18b and WASP-103b, which instead show relatively featureless black-body spectra. The difference can be attributed to the water molecules disassociating on the day-side of ultra-hot Jupiters, and to the presence of opacity owing to H$^{-}$ ions \citep{2018ApJ...855L..30A, 2018AJ....156...17K, 2018arXiv180500096P}.

Hot, fast-rotating stars usually give poor radial-velocity measurements owing to their broad and weak spectral lines. This means that planets around such stars are often confirmed by Doppler tomography of the stellar line profiles through a transit. This method involves detecting the perturbation to stellar line profiles that occurs during transit due to the planet blocking a portion of the Doppler-shifted stellar light.

The first planet discovered in this way was WASP-33b \citep{2010MNRAS.407..507C}, while recently such discoveries include: XO-6b \citep{2017AJ....153...94C}, KELT-17b \citep{2016AJ....152..136Z}, KELT-9b \citep{2017Natur.546..514G}, KELT-19Ab \citep{2018AJ....155...35S}, KELT-20b/MASCARA-2b \citep[][]{2017AJ....154..194L,2018A&A...612A..57T}, KELT-21b \citep{2018AJ....155..100J}, HAT-P-57b \citep{2015AJ....150..197H}, HAT-P-67b \citep{2017AJ....153..211Z}, Kepler-448b \citep{2015A&A...579A..55B}, WASP-167b/KELT-13b \citep{2017MNRAS.471.2743T} and MASCARA-1b \citep{2017A&A...606A..73T}.

We report here the discovery of a hot Jupiter found as a candidate in the WASP-South transit survey \citep{2011EPJWC..1101004H} and confirmed by Doppler tomography using the ESO 3.6-m/HARPS spectrograph \citep{2002Msngr.110....9P}, together with follow-up photometry from the TRAPPIST-South and SPECULOOS-Europa telescopes \citep{2011Msngr.145....2J,2017haex.bookE.130B}. The methods used here are similar to those used for WASP-167b/KELT-13b \citep{2017MNRAS.471.2743T}, but we provide key details of the analysis in Sections ~\ref{sec:specanalysis} to \ref{sec:ageanalysis}.

\section{Observations}
\label{sec:data}
The discovery photometry for \widb\ was obtained using WASP-South, an array of eight cameras based at the South African Astronomical Observatory (SAAO), from 2006 May--2012 June. We used 30-s exposures and typically 10-minute cadence with a 400--700\,nm broad-band filter. WASP-South data are reduced as explained by \citet{2006MNRAS.373..799C} while the candidate selection process is explained by \citet{2007MNRAS.380.1230C}.

Following the detection of a planet-like transit signal with a $\sim$\,4-day period we selected the object for our followup programme.  While the dip is V-shaped, more typical of an eclipsing binary than a planet transit, such dips are also produced by planet transits with a high impact factor. Rejecting eclipsing-binary mimics usually takes only one or two spectra, and so we don't reject V-shaped candidates from WASP follow-up.

We thus obtained 16 radial-velocity measurements using the Euler/CORALIE  spectrograph \citep{2001Msngr.105....1Q}. These were compatible with the transiting object being a planet, however, the broad spectral features meant that the error bars are large and thus could not produce a secure orbital variation and hence a mass. To confirm the planet we therefore decided to also use Doppler tomography, and observed a series of 23 spectra with the HARPS spectrograph covering a transit on the night of Mar 13th 2016. Simultaneously with this we observed the transit photometrically with TRAPPIST-South. Details of the observations are given in Table~\ref{table:observations} while the measured radial velocities are given in Table ~\ref{table:RVs}.

We have also obtained photometry of three other transits with TRAPPIST-South and SPECULOOS (see Table~\ref{table:observations}). While TRAPPIST-South has been used extensively for the discovery and parametrisation of WASP planets \citep{2012A&A...542A...4G}, this is the first WASP paper to feature data from the newer SPECULOOS, so we describe it briefly.

The SPECULOOS-Europa telescope is one of four identical telescopes currently being installed at ESO Paranal Observatory. SPECULOOS is a ground-based transit survey that will search for Earth-sized planets transiting the nearest ultracool dwarfs (\citet{2017haex.bookE.130B}).  Each SPECULOOS telescope is a robotic Ritchey-Chretien (F/8) telescope of 1-m diameter. They are equipped with Andor Peltier-cooled deeply depleted 2K\,$\times$\,2K CCD cameras, with 13.5 micron pixels. The field of view of each telescope is 12'\,$\times$\,12' and the corresponding pixel scale is 0.35''\,pixel$^{-1}$. 

Lastly, we report that we searched the WASP photometry looking for stellar rotational modulations in the range 0--1.5 cycles day$^{-1}$, using the methods of \citet{2011PASP..123..547M}. We did not detect any modulations, or evidence of pulsations, with an upper limit of 0.8\,mmag.

\begin{table}
\caption{Details of all observations of \widb\ used in this work, including the discovery photometry, the follow-up photometry and the spectroscopic observations.}
\centering
\begin{tabular}{lcc}
\hline
Facility & Date & Notes \\[0.5ex]
\hline
WASP-South & 2006-05-- & 35883 points \\
 & 2012-06 & \\
TRAPPIST-South & 2014-03-20 & I+z'. 14s exp. \\
TRAPPIST-South & 2016-03-13 & I+z'. 8s exp. \\
TRAPPIST-South & 2017-03-08 & V. 15s exp. \\
SPECULOOS-Europa & 2017-07-13 & I+z'. 10s exp. \\
CORALIE	& 2014-03--  & 16 out-of-transit \\
 & 2017-08 & spectra \\
HARPS & 2016-03-13 & 23 spectra taken \\ 
 & & including a transit \\ [1ex]
\hline
\end{tabular}
\label{table:observations}
\end{table}

\begin{table}
\caption{Radial velocities and bisector spans for \widb .}
\centering
\begin{tabular}{lrrrr}
\hline
BJD (TDB & RV & $\sigma$$_{\rm RV}$ & BS & $\sigma$$_{\rm BS}$ \\
--2,450,000)  & (km s$^{-1}$) & (km s$^{-1}$) & (km s$^{-1}$) & (km s$^{-1}$) \\ [0.5mm]
\hline
\multicolumn{5}{l}{CORALIE RVs:} \\
6719.750940 & 4.87 & 0.05 & --0.28 & 0.10 \\
6770.634386 & 4.91 & 0.05 & --0.15 & 0.10 \\
6836.575290 & 4.84 & 0.09 & 0.11 & 0.18 \\
7072.738390 & 4.71 & 0.06 & 0.17 & 0.12 \\
7888.597863	& 4.79 & 0.09 & --0.21 & 0.18 \\
7890.515926	& 4.73 & 0.14 & 0.24 & 0.28 \\
7894.502532	& 4.79 & 0.08 & --0.10 & 0.16 \\
7903.605663	& 4.66 & 0.13 & --0.42 & 0.26 \\
7905.689111	& 4.85 & 0.07 & --0.30 & 0.14 \\
7917.567974	& 4.85 & 0.07 & --0.06 & 0.14 \\
7924.505661	& 4.82 & 0.06 & --0.35 & 0.12 \\
7951.506527	& 4.73 & 0.17 & --0.41 & 0.34 \\
7954.495100	& 4.83 & 0.07 & 0.01 & 0.14 \\
7959.517031	& 4.70 & 0.09 & --0.03 & 0.18 \\
7973.492569	& 4.82 & 0.12 & 0.17 & 0.24 \\
7974.518841 \medskip & 4.65 & 0.09 & --0.16 & 0.18 \\
\multicolumn{5}{l}{HARPS RVs:} \\
7461.571827 & 4.87 & 0.02 & --0.09 & 0.04 \\
7461.582499 & 4.89 & 0.02 & --0.16 & 0.04 \\
7461.593380 & 4.88 & 0.02 & --0.09 & 0.04 \\
7461.604248 & 4.85 & 0.02 & --0.16 & 0.04 \\
7461.615140 & 4.89 & 0.02 & --0.15 & 0.04 \\
7461.625708 & 4.86 & 0.02 & --0.10 & 0.04 \\
7461.636692 & 4.85 & 0.02 & --0.12 & 0.04 \\
7461.647260 & 4.89 & 0.02 & --0.10 & 0.04 \\
7461.657920 & 4.86 & 0.02 & --0.03 & 0.04 \\
7461.668696 & 4.88 & 0.02 & --0.14 & 0.04 \\
7461.679576 & 4.83 & 0.02 & --0.09 & 0.04 \\
7461.690352 & 4.78 & 0.02 & 0.12 & 0.04 \\
7461.701024 & 4.77 & 0.01 & 0.07 & 0.02 \\
7461.711800 & 4.79 & 0.01 & --0.03 & 0.02 \\
7461.722576 & 4.78 & 0.02 & --0.13 & 0.04 \\
7461.733457 & 4.81 & 0.01 & --0.15 & 0.02 \\
7461.743804 & 4.83 & 0.02 & --0.10 & 0.04 \\
7461.754893 & 4.84 & 0.02 & --0.00 & 0.04 \\
7461.765565 & 4.88 & 0.02 & --0.10 & 0.04 \\
7461.776549 & 4.85 & 0.02 & --0.19 & 0.04 \\
7461.787013 & 4.87 & 0.02 & --0.15 & 0.04 \\
7461.797985 & 4.87 & 0.02 & --0.02 & 0.04 \\
7461.808866 \medskip & 4.84 & 0.02 & --0.19 & 0.04 \\
\\
\hline
\end{tabular}
\label{table:RVs}
\end{table}

\section{Spectral Analysis}
\label{sec:specanalysis}
We first performed a spectral analysis on a median-stacked HARPS spectrum created from the 23 we obtained, in order to determine some stellar properties. We follow the method described by \citet{2013MNRAS.428.3164D} to determine values for the stellar effective temperature \teff , stellar surface gravity $\log{g_{*}}$, the stellar metallicity $\rm [Fe/H]$, the stellar lithium abundance $\rm log\,A(Li)$ and the projected stellar rotational velocity $ v \sin i_\star$. To constrain the latter we obtain a macroturbulence value of $v_{\rm mac}$\,=\,6.3 km\,s$^{-1}$ using the \citet{2014MNRAS.444.3592D} calibration.  \teff\ was measured using the H$\alpha$ line while $\log{g_{*}}$ was measured from the Na D lines. We also determine the spectral type of the star to be F6V, by using the MKCLASS program \citep{2014AJ....147...80G}. The values obtained for each of the fitted parameters are given in Table~\ref{table:allResults}.

\section{Combined Analyses}
\label{sec:combanalysis}
We performed a Markov Chain Monte Carlo (MCMC) fitting procedure which uses the stellar parameters obtained in the spectral analysis (Section~\ref{sec:specanalysis}) to constrain the fit. We used the latest version of the MCMC code described by \citet{2007MNRAS.380.1230C} and \citet{2008MNRAS.385.1576P}, which is capable of fitting photometric, RV and tomographic data simultaneously \citep{2010MNRAS.403..151C}. 

The system parameters which are determined from the photometric data are the epoch of mid-transit $T_{\rm c}$, the orbital period $P$, the planet-to-star area ratio $(R_{\rm p}/R_{\star})^{2}$, the transit duration $T_{\rm 14}$, and the impact parameter $b$. Limb darkening was accounted for using the \citet{2000A&A...363.1081C, 2004A&A...428.1001C} four-parameter non-linear law: for each new value of \teff\ a set of parameters is interpolated from the Claret tables. The proposed values of the stellar mass are calculated using the Enoch--Torres relation \citep{2010A&A...516A..33E, 2010A&ARv..18...67T}.

The RV fitting then provides values for the stellar reflex velocity semi-amplitude $K_{\rm 1}$ and the barycentric system velocity $\gamma$. We assume a circular orbit, since we do not have sufficient quality in the out-of-transit RVs to constrain the eccentricity. In any case, hot Jupiters often settle into circular orbits on time-scales that are shorter than their lifetimes through tidal circularization \citep{2011MNRAS.414.1278P}, so usually their orbits are circular. If there are accurate RVs taken through transit, it is also possible to measure the projected spin-orbit misalignment angle $\lambda$ by fitting the RM effect.

The 23 HARPS spectra were cross-correlated using the standard HARPS Data Reduction Software over a window of $\pm$ 350 km s$^{-1}$ (as described in \citet{1996A&AS..119..373B}, \citet{2002Msngr.110....9P}). The cross-correlation functions (CCFs) were created using a mask matching a G2 spectral type, containing zeroes at the positions of absorption lines and ones in the continuum. The tomographic data are then comprised of the time series of CCFs taken through transit. The CORALIE spectra were also correlated using the same methodology.

We used the MCMC code in two modes.  The first mode fits the CCFs to obtain RV values, and then uses the calibrations of \citet{2011ApJ...742...69H} to model the RM effect and thus measure  $\lambda$. The second mode fits the in-transit CCFs directly, modelling the perturbations of the CCFs due to the path of the planet across the stellar disc \citep[e.g.][]{2017MNRAS.464..810B,2017MNRAS.471.2743T}. The parameters determined in this part of the analysis are $v \sin i_\star$, $\lambda$, the stellar line-profile Full-Width at Half-Maximum (FWHM), the FWHM of the line perturbation due to the planet $v_{\rm FWHM}$ and the system $\gamma$-velocity. The MCMC code assumes a Gaussian shape for the line perturbation caused by the planet. We obtain initial values for the stellar line FWHM and the $\gamma$-velocity by fitting a Gaussian profile to the CCFs and apply the spectral $v \sin i_\star$ and \teff\ as priors. Neither $\lambda$ nor $v_{\rm FWHM}$ had a prior applied.

We give the solutions obtained using the two modes in Table ~\ref{table:allResults}. Both fits gave strongly consistent results. We adopt the solution of the fit including tomography, since it is a more direct method that uses more of the line-profile information. 

\subsection{A grazing transit}
\begin{figure}
\hspace*{2mm}\includegraphics[width=0.49\textwidth]{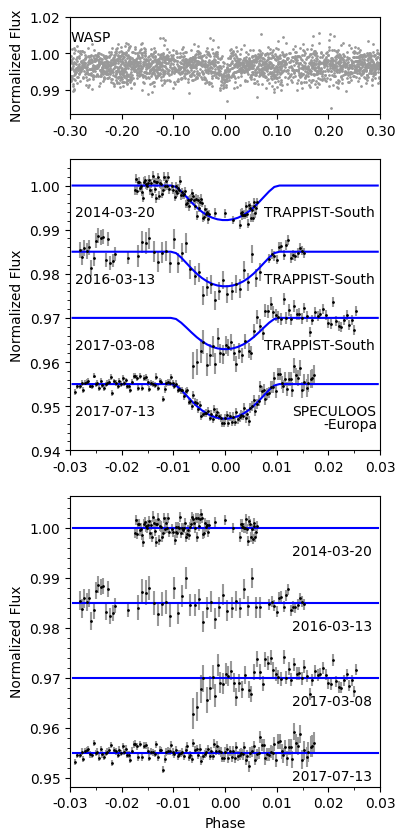}\\ [-2mm]
\caption{The WASP discovery photometry (top) and follow-up transit lightcurves (middle). The blue lines show the final model obtained in the MCMC fitting (see Section~\ref{sec:combanalysis}). The bottom panel then shows the residuals of the fit.}
\label{fig:phot}
\end{figure}

\begin{figure}
\hspace*{2mm}\includegraphics[width=0.49\textwidth]{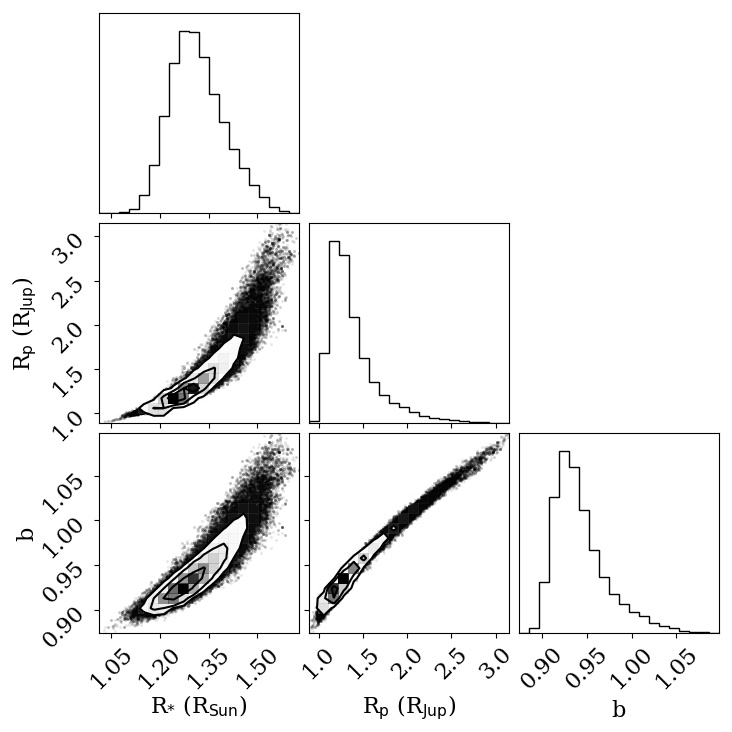}\\ [-2mm]
\caption{Probability distributions for the parameters $R_{\rm p}$, $R_{*}$ and $b$, created from the results of the second-mode MCMC run.}
\label{fig:probability}
\end{figure}

The photometry and the best-fitting model are shown in Fig.~\ref{fig:phot}. We found that constraining the photometric fit was difficult since the transit is either grazing or near-grazing and does not show clear 2$^{\rm nd}$ and 3$^{\rm rd}$ contacts. This means that $R_{\rm p}$/$R_{*}$ and the impact parameter $b$ are poorly constrained. We show the probability distributions of $R_{\rm p}$,$R_{*}$ and $b$ in Fig.~\ref{fig:probability}.

We calculated the ``grazing criterion'', namely ($R_{\rm p}/R_{*} + b$), which if >\,1 implies a grazing transit \citep{2011A&A...526A.130S}. We obtain 1.02$^{+ 0.04}_{-0.02}$, which means that we cannot securely distinguish between grazing and near-grazing solutions.

We used the InfraRed Flux Method \citep[IRFM, ][]{1977MNRAS.180..177B} to obtain values for \teff\ and the angular diameter $\theta$ of \wid\ , which are quoted in Table~\ref{table:allResults}. We then used $\theta$ and the {\it Gaia} DR2 \citep{2016A&A...595A...1G,2018arXiv180409365G} parallax, which is also quoted in Table~\ref{table:allResults}, to estimate the stellar radius. We took reddening into account by measuring the equivalent width of the interstellar Na D lines using the stacked HARPS spectrum from Section~\ref{sec:specanalysis}, finding a width of 80\,m$\AA$ which equates to an extinction value of $E(B-V)$\,=\,0.02 \citep{1997A&A...318..269M}. We have also taken into account the systematic offset in the {\it Gaia} parallax value (of 0.082\,mas), as measured by \citet{2018ApJ...862...61S}. We obtain a stellar radius of 1.35\,$\pm$\,0.10\,$R_{\rm \odot}$ which is consistent with our fitted radius of 1.31 $\pm$ 0.08\,$R_{\rm \odot}$.

\subsection{The planet's mass}
\begin{figure}
\hspace*{2mm}\includegraphics[width=0.46\textwidth]{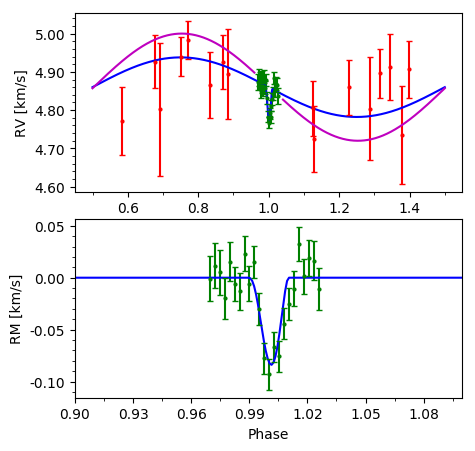}\\ [-2mm]
\caption{Top: The 16 CORALIE RVs (red points) obtained for \widb\ . The magenta line shows the expected RV amplitude for a planet of 1.3\,M$_{\rm Jup}$, our derived upper limit (95\,\% confidence). The blue line shows the best-fit model including the RM fit. Bottom: the 23 through-transit HARPS RVs (green points). The blue line shows the best-fit model with the Keplerian RV curve subtracted, leaving only the fit to the RM effect.}
\label{fig:RV-RM}
\end{figure}

The CORALIE and HARPS RVs are shown in Fig.~\ref{fig:RV-RM}. Due to the relatively large error bars in the out-of-transit RV measurements we do not regard the fitted semi-amplitude (of 0.08\,$\pm$\,0.03\,km\,s$^{-1}$) to be a measure of the planet's mass. However, we were able to put a 95\,\% confidence upper limit on the mass of 1.3\,M$_{\rm Jup}$, and the predicted curve for this value is also shown in Fig.~\ref{fig:RV-RM}.

\subsection{The Doppler track}

\begin{figure*}
\centering
\hspace*{2mm}\includegraphics[width=0.9\textwidth]{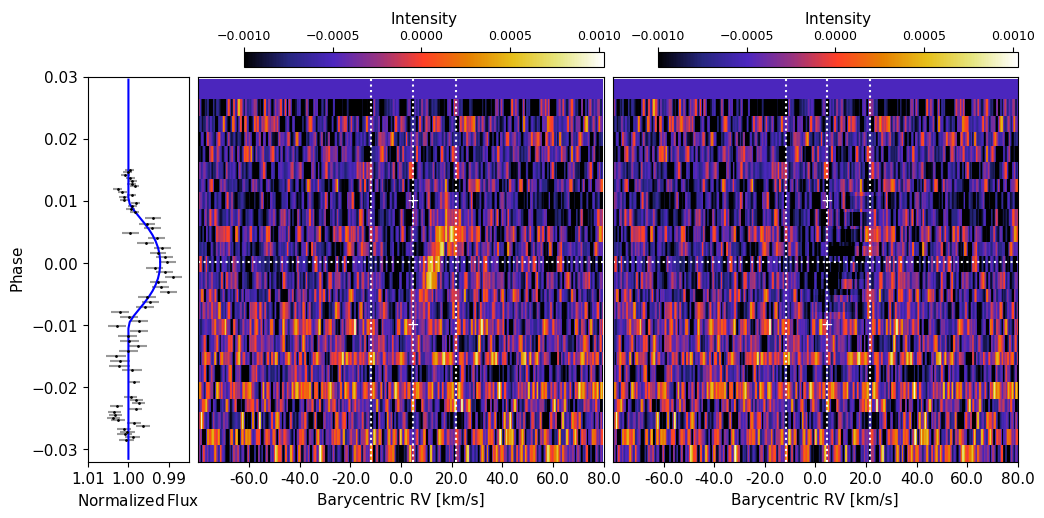}\\ [-2mm]
\caption{Middle: The line profiles through transit, with the average of the out-of-transit CCFs subtracted. We interpret this tomogram as showing a prograde-moving planet signal in the red-shifted section of the tomogram. Right: the line profile residuals after subtracting the planet model. In both of these panels, the white dotted vertical lines mark the positions of the $\gamma$ velocity of the system and the positions of $\gamma \pm v \sin i_{\star}$. The phase of mid-transit is marked by the white horizontal dotted line. The white + symbols indicate the beginning and end of the transit event, calculated using the ephemeris obtained in the adopted solution. Left: The TRAPPIST-South lightcurve taken simultaneously with the tomographic observation.}
\label{fig:SW1303tomog3}
\end{figure*}

We display the tomographic data as a function of the planet's orbital phase in Fig.~\ref{fig:SW1303tomog3}. In creating this plot we first remove the invariant stellar line profile by subtracting the average of the out-of-transit CCFs. We also display the simultaneous photometric observation to the left of the tomogram, and the residuals from subtracting the planet model on the right.

\begin{figure}
\hspace*{2mm}\includegraphics[width=0.4\textwidth]{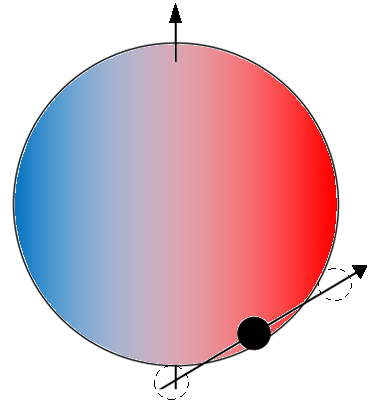}\\ [-2mm]
\centering
\caption{The transit chord calculated from the fitted values of $R_{\rm p}$, $R_{*}$, $b$ and $\lambda$ (see Table ~\ref{table:allResults}.) The dashed circles show the positions of the planet at 1$^{\rm st}$ and 4$^{\rm th}$ contacts.}
\label{fig:chord}
\end{figure}

We interpret the resulting tomogram as showing a faint, prograde-moving planet signal crossing only the red-shifted portion of the plot.  This is in line with the transit being grazing, such that the planet crosses only a short chord on the face of the star (see Fig.~\ref{fig:chord}). 

The planet's Doppler shadow appears very faint  at the beginning and end of the transit (see Fig.~\ref{fig:SW1303tomog3}). This is likely due to there being little of the planet on the face of the star near 1$^{\rm st}$ and 4$^{\rm th}$ contacts, owing to the near-grazing nature of the orbit.

\section{Stellar Age Determination}
\label{sec:ageanalysis}
\begin{figure}
\centering
\hspace*{2mm}\includegraphics[width=0.49\textwidth]{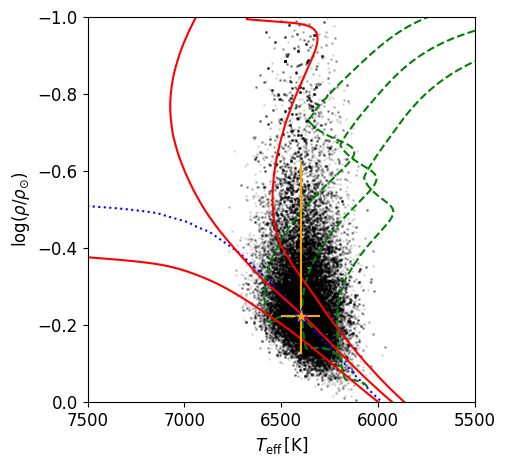}\\ [-2mm]
\caption{The best fitting evolutionary tracks and isochrones of \wid\ obtained using {\sc bagemass}. Black points: individual steps in the MCMC. Dotted blue line: Zero-Age Main Sequence (ZAMS) at best-fit [Fe/H]. Green dashed lines: evolutionary track for the best-fit [Fe/H] and mass, plus $1\sigma$ bounds. Red lines: isochrone for the best-fit [Fe/H] and age, plus $1\sigma$ bounds. Orange star: measured values of \teff\ and $\rho_{*}$ for \wid\ obtained in the spectral and photometric analyses respectively.}
\label{fig:trho_plot}
\end{figure}

We estimated the age of \wid\ using the open source software {\sc bagemass}\footnote{\url{http://sourceforge.net/projects/bagemass}}.   {\sc bagemass} uses the Bayesian method of \citet{2015A&A...575A..36M} to fit the age, mass and initial metallicity of a star using the {\sc garstec} stellar evolution code \citep{2008Ap&SS.316...99W}. We applied constraints on the stellar temperature and metallicity (\teff\,=\,6400\,$\pm$\,100\,K and [Fe/H]\,=\,0.09\,$\pm$\,0.09 as obtained in the spectral analysis) as well as the stellar density ($\rho_{*}/\rho_{\odot}$\,=\,0.6\,$\pm$\,0.2 from the transit analysis). We adopt the solution obtained for a solar mixing length and He abundance, since enhancing the He abundance made no significant change to the fit while reducing the solar mixing length worsened the fit. We display the resulting isochrones and evolutionary tracks for this fit in Fig.~\ref{fig:trho_plot} and the fitted values are given in Table ~\ref{table:allResults}.

We find \wid\ to be consistent with a main-sequence star or one beginning to evolve off the main sequence. The Li abundance obtained in Section~\ref{sec:specanalysis} is also consistent with the star being non-evolved, but for mid-F stars the Li abundance is not a good age indicator. For the measured value of $\log{A(Li)}$\,=\,2.48\,$\pm$\,0.10, \wid\ could be up to a few Gyr old \citep{2005A&A...442..615S}. If we define the main-sequence lifetime of a star to be the time taken for all hydrogen in the core to be exhausted, we can use the best-fit evolutionary track from {\sc bagemass} to estimate the age at which \wid\ will leave the main sequence: 4.3\,$\pm$\,0.6\,Gyr.

\begin{table} 
\caption{All system parameters obtained for \widb\ in this work.} 
\label{table:allResults}
\centering
\begin{tabular}{lcc}
\hline
\multicolumn{3}{l}{1SWASP\,J130310.57--412305.3}\\
\multicolumn{3}{l}{2MASS\,J13031055--4123053}\\
\multicolumn{3}{l}{TIC ID:102192004}\\
\multicolumn{3}{l}{RA\,=\,13$^{\rm h}$03$^{\rm m}$10.57$^{\rm s}$, 
Dec\,=\,--41$^{\circ}$23$^{'}$05.3$^{''}$ (J2000)}\\
\multicolumn{3}{l}{$V$ = 11.9 (NOMAD)}  \\
\multicolumn{3}{l}{IRFM \teff\ = 6380 $\pm$ 140 K}  \\ 
\multicolumn{3}{l}{IRFM $\theta$ = 0.031 $\pm$ 0.002 mas}  \\
\multicolumn{3}{l}{{\it Gaia} DR2 Proper Motions:} \\
\multicolumn{3}{l}{(RA) 0.043\,$\pm$\,0.071 (Dec) --5.784\,$\pm$\,0.112 mas/yr} \\
\multicolumn{3}{l}{{\it Gaia} DR2 Parallax: 2.41\,$\pm$\,0.06\,mas} \\
\multicolumn{3}{l}{Rotational Modulations: < 0.8 mmag (95\%)}\\
\hline
\multicolumn{3}{l}{\it{Stellar parameters from spectral analysis:}} \\[0.5ex]
Parameter & \multicolumn{2}{c}{Value} \\
(Unit) & & \\
Spectral type & \multicolumn{2}{c}{F6V} \\
\teff\ (K) & \multicolumn{2}{c}{6400\,$\pm$\,100} \\
$\log g_{*}$ & \multicolumn{2}{c}{4.15\,$\pm$\,0.15} \\
{[Fe/H]} & \multicolumn{2}{c}{0.09\,$\pm$\,0.09} \\
$\log{A(Li)}$ & \multicolumn{2}{c}{2.48\,$\pm$\,0.10} \\
$v \sin i_{\rm *}$ (km\,s$^{-1}$) & \multicolumn{2}{c}{$16.5\,\pm\,0.5$} \\
$v_{\rm mac}$ (km\,s$^{-1}$)  \medskip & \multicolumn{2}{c}{6.3\,km\,s$^{-1}$} \\
\multicolumn{3}{l}{\it{Parameters from photometric and RV analysis:}} \\[0.5ex]
Parameter & DT Value & RM Value: \\
(Unit) & (adopted): & \\
$P$ (d) & 4.233700 $\pm$ 0.000003 & 4.233700 $\pm$ 0.000003 \\
$T_{\rm c}$ (BJD$_{\rm TDB}$) & 2457465.9336 $\pm$ 0.0004 & 2457465.9335 $\pm$ 0.0004 \\
$T_{\rm 14}$ (d) & 0.085 $\pm$ 0.002 & 0.085 $\pm$ 0.002 \\
$R_{\rm P}^{2}$/R$_{*}^{2}$ & 0.0086 $\pm$ 0.0003 & 0.0088 $\pm$ 0.0006 \\
$b$ & 0.94 $\pm$ 0.03 & 0.95 $\pm$ 0.04 \\
$i$ ($^\circ$) & 84.2 $\pm$ 0.5 & 84.0 $\pm$ 0.7 \\
$a$ (AU)  & 0.0559 $\pm$ 0.0009 & 0.0555 $\pm$ 0.0009 \\
$M_{\rm *}$ ($M_{\rm \odot}$) & 1.30 $\pm$ 0.06 & 1.31 $\pm$ 0.07 \\
$R_{\rm *}$ ($R_{\rm \odot}$) & 1.31 $\pm$ 0.08 & 1.3 $\pm$ 0.1 \\
$\log g_{*}$ (cgs) & 4.32 $\pm$ 0.04 & 4.31 $\pm$ 0.06 \\
$\rho_{\rm *}$ ($\rho_{\rm \odot}$) & 0.6 $\pm$ 0.2 & 0.6 $\pm$ 0.1 \\
$T_{\rm eff}$ (K) & 6400 $\pm$ 100 & 6400 $\pm$ 100 \\
{[Fe/H]} & 0.09 $\pm$ 0.09 & 0.09 $\pm$ 0.09 \\
$M_{\rm P}$ ($M_{\rm Jup}$) & <\,1.3 (95\%) & <\,1.3 (95\%) \\
$K$ (km s$^{-1}$) & <\,0.14 (95\%) & <\,0.14 (95\%) \\
$R_{\rm P}$ ($R_{\rm Jup}$) & 1.3 $\pm$ 0.5 & 1.4 $\pm$ 0.5 \\
$T_{\rm eql}$ (K) \medskip & 1490 $\pm$ 50 & 1500 $\pm$ 60 \\
\multicolumn{3}{l}{\it{Parameters from RM and DT analyses:}}\\[0.5ex]
$\gamma$ (km s$^{-1}$)& 4.864 $\pm$ 0.005 & 4.860 $\pm$ 0.004 \\
$\lambda$ ($^\circ$) \medskip & 31 $\pm$ 1 & 34 $\pm$ 5 \\
\multicolumn{3}{l}{\it{Parameters from} {\sc bagemass}:} \\[0.5ex]
Parameter & \multicolumn{2}{c}{Value} \\
(Unit) & & \\
Age (Gyr) & \multicolumn{2}{c}{1.65\,$\pm$\,0.85} \\[0.5ex]
$M_{*}$ ($M_{\rm \odot}$) & \multicolumn{2}{c}{1.28\,$\pm$\,0.07} \\
$\rm{[Fe/H]}_{\rm init}$ \medskip & \multicolumn{2}{c}{0.12\,$\pm$\,0.08} \\
\hline
\end{tabular} 
\end{table}

\section{Discussion and Conclusions}
\widb\ is revealed by Doppler tomography to be a planet making a grazing transit of its host star in a misaligned orbit with an alignment angle of  $\lambda$\,=\,31$^\circ$\,$\pm$\,1$^\circ$. 

\wid\ is an F6 star with an effective temperature of \teff = 6400 $\pm$ 100 K and a measured $v \sin i_\star$ of 16.5\,$\pm$\,0.5\,km\,s$^{-1}$. This rotation rate, together with a fitted radius of 1.31\,$\pm$\,0.08\,R$_{\odot}$, implies a stellar rotation period of \prot\,<\,4.4 d. Since the planet's orbital period is 4.23 d, this means that the stellar rotation period could be, but is not certain to be, shorter than the planet's orbit.  Most hot-Jupiter systems have rotation periods that are longer than the orbit, but having  \porb\ > \prot\ has been found for other hot, more rapidly rotating host stars, including KELT-17b \citep{2016AJ....152..136Z}, WASP-167b/KELT-13b \citep{2017MNRAS.471.2743T} and XO-6b \citep{2017AJ....153...94C}.  In systems with \porb\ < \prot\ and with prograde orbits the tidal interaction is thought to produce decay of the planet's orbit, but this will be reversed in systems such as \wid, with a prograde orbit and with \porb > \prot\ (see the discussions in \citet{2017AJ....153...94C} and \citet{2017MNRAS.471.2743T}). The difference in dynamical evolution of hot-star hot Jupiters makes them interesting targets and is one reason for finding more examples of such systems.

Another dynamical difference is that hot-Jupiter orbits are much more likely to be misaligned around hotter stars, which might be related to reduced tidal damping in hotter stars with smaller or absent convective envelopes \citep{2010ApJ...718L.145W}. With a misaligned orbit \widb\ is in line with this trend. Of the 12 other systems confirmed with tomographic methods, 8 are at least moderately misaligned. These are WASP-33b \citep{2010MNRAS.407..507C}, HAT-P-57b \citep{2015AJ....150..197H}, KELT-17b \citep{2016AJ....152..136Z}, KELT-9b \citep{2017Natur.546..514G}, KELT-19Ab \citep{2018AJ....155...35S}, XO-6b \citep{2017AJ....153...94C}, WASP-167b/KELT-13b \citep{2017MNRAS.471.2743T} and MASCARA-1b \citep{2017A&A...606A..73T}. 

High stellar irradiation produces hotter planetary atmospheres, and is thought to result in the inflated radii seen in many hot Jupiters \citep[e.g.][]{2016AJ....152..182H,2017AJ....153..211Z,2018AJ....155...35S}. With an equilibrium temperature of 1490\,$\pm$\,50\,K, we would thus expect \widb\ to be moderately inflated. 

The actual planetary radius is hard to measure owing to the grazing or near-grazing transit, which means that 2$^{\rm nd}$ and 3$^{\rm rd}$  contacts are not visible in the transit profile and the fitted radius is degenerate with the impact parameter (Fig.~\ref{fig:probability}). Thus we can do no better than loosely constraining the radius to $R_{P} = 1.3 \pm 0.5$ R$_{\rm Jup}$, which is consistent with that of an inflated hot Jupiter.

The mass of \widb\ is also uncertain, since the hot host star limits the accuracy and precision of radial-velocity measurements. We report only an upper limit of 1.3\,M$_{\rm Jup}$, so again \widb\ is most likely a fairly typical inflated hot Jupiter. It may be possible, however, to constrain the mass further with some more precise RV measurements taken out-of-transit using HARPS.

At $V$ = 11.9, \wid\ is the faintest hot-Jupiter system for which the shadow of the planet has been detected by tomographic methods.  The next faintest are Kepler-448 at $V$ = 11.4 \citep{2015A&A...579A..55B} and HAT-P-56 at $V$ = 10.9 \citep{2015AJ....150...85H,2016MNRAS.460.3376Z}, which was initially confirmed with radial velocity measurements.  

HAT-P-56b is also comparable in that it has a near-grazing transit with an impact parameter of $b = 0.873^{+0.004}_{-0.006}$ \citep{2015AJ....150...85H}, which compares with $b = 0.94 \pm 0.03$ for \widb\ . As with our work the tomographic planet trace for HAT-P-56b is faint and possibly shows evidence for getting fainter when the planet is only partially occulting the star \citep[i.e. at the beginning and end of the transit, ][]{2016MNRAS.460.3376Z}.

\section*{Acknowledgements}
WASP-South is hosted by the South African Astronomical Observatory and we are grateful for their ongoing support and assistance. Funding for WASP comes from consortium universities and from the UK's Science and Technology Facilities Council. The Euler Swiss telescope is supported by the Swiss National Science Foundation. TRAPPIST-South is funded by the Belgian Fund for Scientific Research (Fond National de la Recherche Scientifique, FNRS) under the grant FRFC 2.5.594.09.F, with the participation of the Swiss National Science Foundation (SNF). We acknowledge use of the ESO 3.6-m/HARPS under program 096.C-0762. MG is FNRS Research Associate, and EJ is FNRS Senior Research Associate. The research leading to these results has received funding from the European Research Council under the FP/2007-2013 ERC Grant Agreement n° 336480, and from the ARC grant for Concerted Research Actions, financed by the Wallonia-Brussels Federation. This work was also partially supported by a grant from the Simons Foundation (ID 327127 to  Didier Queloz).


\bibliographystyle{mnras}
\bibliography{litbiblio}

\begin{thebibliography}{}
\makeatletter
\relax
\def\mn@urlcharsother{\let\do\@makeother \do\$\do\&\do\#\do\^\do\_\do\%\do\~}
\def\mn@doi{\begingroup\mn@urlcharsother \@ifnextchar [ {\mn@doi@}
  {\mn@doi@[]}}
\def\mn@doi@[#1]#2{\def\@tempa{#1}\ifx\@tempa\@empty \href
  {http://dx.doi.org/#2} {doi:#2}\else \href {http://dx.doi.org/#2} {#1}\fi
  \endgroup}
\def\mn@eprint#1#2{\mn@eprint@#1:#2::\@nil}
\def\mn@eprint@arXiv#1{\href {http://arxiv.org/abs/#1} {{\tt arXiv:#1}}}
\def\mn@eprint@dblp#1{\href {http://dblp.uni-trier.de/rec/bibtex/#1.xml}
  {dblp:#1}}
\def\mn@eprint@#1:#2:#3:#4\@nil{\def\@tempa {#1}\def\@tempb {#2}\def\@tempc
  {#3}\ifx \@tempc \@empty \let \@tempc \@tempb \let \@tempb \@tempa \fi \ifx
  \@tempb \@empty \def\@tempb {arXiv}\fi \@ifundefined
  {mn@eprint@\@tempb}{\@tempb:\@tempc}{\expandafter \expandafter \csname
  mn@eprint@\@tempb\endcsname \expandafter{\@tempc}}}

\bibitem[\protect\citeauthoryear{{Arcangeli} et~al.,}{{Arcangeli}
  et~al.}{2018}]{2018ApJ...855L..30A}
{Arcangeli} J.,  et~al., 2018, \mn@doi [\apjl] {10.3847/2041-8213/aab272},
  \href {http://adsabs.harvard.edu/abs/2018ApJ...855L..30A} {855, L30}

\bibitem[\protect\citeauthoryear{{Baranne} et~al.,}{{Baranne}
  et~al.}{1996}]{1996A&AS..119..373B}
{Baranne} A.,  et~al., 1996, \aaps, \href
  {http://adsabs.harvard.edu/abs/1996A%26AS..119..373B} {119, 373}

\bibitem[\protect\citeauthoryear{{Beatty}, {Madhusudhan}, {Tsiaras}, {Zhao},
  {Gilliland}, {Knutson}, {Shporer}  \& {Wright}}{{Beatty}
  et~al.}{2017}]{2017AJ....154..158B}
{Beatty} T.~G.,  {Madhusudhan} N.,  {Tsiaras} A.,  {Zhao} M.,  {Gilliland}
  R.~L.,  {Knutson} H.~A.,  {Shporer} A.,   {Wright} J.~T.,  2017, \mn@doi
  [\aj] {10.3847/1538-3881/aa899b}, \href
  {http://adsabs.harvard.edu/abs/2017AJ....154..158B} {154, 158}

\bibitem[\protect\citeauthoryear{{Blackwell} \& {Shallis}}{{Blackwell} \&
  {Shallis}}{1977}]{1977MNRAS.180..177B}
{Blackwell} D.~E.,  {Shallis} M.~J.,  1977, \mn@doi [\mnras]
  {10.1093/mnras/180.2.177}, \href
  {http://adsabs.harvard.edu/abs/1977MNRAS.180..177B} {180, 177}

\bibitem[\protect\citeauthoryear{{Bourrier} et~al.,}{{Bourrier}
  et~al.}{2015}]{2015A&A...579A..55B}
{Bourrier} V.,  et~al., 2015, \mn@doi [\aap] {10.1051/0004-6361/201525750},
  \href {http://adsabs.harvard.edu/abs/2015A%26A...579A..55B} {579, A55}

\bibitem[\protect\citeauthoryear{{Brown} et~al.,}{{Brown}
  et~al.}{2017}]{2017MNRAS.464..810B}
{Brown} D.~J.~A.,  et~al., 2017, \mn@doi [\mnras] {10.1093/mnras/stw2316},
  \href {http://adsabs.harvard.edu/abs/2017MNRAS.464..810B} {464, 810}

\bibitem[\protect\citeauthoryear{{Burdanov}, {Delrez}, {Gillon}, {Jehin},
  {Speculoos}  \& {Trappist Teams}}{{Burdanov}
  et~al.}{2017}]{2017haex.bookE.130B}
{Burdanov} A.,  {Delrez} L.,  {Gillon} M.,  {Jehin} E.,  {Speculoos} T.,
  {Trappist Teams} 2017, {SPECULOOS Exoplanet Search and Its Prototype on
  TRAPPIST}.
p.~130, \mn@doi{10.1007/978-3-319-30648-3_130-1}

\bibitem[\protect\citeauthoryear{{Claret}}{{Claret}}{2000}]{2000A&A...363.1081C}
{Claret} A.,  2000, \aap, \href
  {http://adsabs.harvard.edu/abs/2000A%26A...363.1081C} {363, 1081}

\bibitem[\protect\citeauthoryear{{Claret}}{{Claret}}{2004}]{2004A&A...428.1001C}
{Claret} A.,  2004, \mn@doi [\aap] {10.1051/0004-6361:20041673}, \href
  {http://adsabs.harvard.edu/abs/2004A%26A...428.1001C} {428, 1001}

\bibitem[\protect\citeauthoryear{{Collier Cameron} et~al.,}{{Collier Cameron}
  et~al.}{2006}]{2006MNRAS.373..799C}
{Collier Cameron} A.,  et~al., 2006, \mn@doi [\mnras]
  {10.1111/j.1365-2966.2006.11074.x}, \href
  {http://adsabs.harvard.edu/abs/2006MNRAS.373..799C} {373, 799}

\bibitem[\protect\citeauthoryear{{Collier Cameron} et~al.,}{{Collier Cameron}
  et~al.}{2007}]{2007MNRAS.380.1230C}
{Collier Cameron} A.,  et~al., 2007, \mn@doi [\mnras]
  {10.1111/j.1365-2966.2007.12195.x}, \href
  {http://adsabs.harvard.edu/abs/2007MNRAS.380.1230C} {380, 1230}

\bibitem[\protect\citeauthoryear{{Collier Cameron}, {Bruce}, {Miller}, {Triaud}
   \& {Queloz}}{{Collier Cameron} et~al.}{2010a}]{2010MNRAS.403..151C}
{Collier Cameron} A.,  {Bruce} V.~A.,  {Miller} G.~R.~M.,  {Triaud}
  A.~H.~M.~J.,   {Queloz} D.,  2010a, \mn@doi [\mnras]
  {10.1111/j.1365-2966.2009.16131.x}, \href
  {http://adsabs.harvard.edu/abs/2010MNRAS.403..151C} {403, 151}

\bibitem[\protect\citeauthoryear{{Collier Cameron} et~al.,}{{Collier Cameron}
  et~al.}{2010b}]{2010MNRAS.407..507C}
{Collier Cameron} A.,  et~al., 2010b, \mn@doi [\mnras]
  {10.1111/j.1365-2966.2010.16922.x}, \href
  {http://adsabs.harvard.edu/abs/2010MNRAS.407..507C} {407, 507}

\bibitem[\protect\citeauthoryear{{Crouzet} et~al.,}{{Crouzet}
  et~al.}{2017}]{2017AJ....153...94C}
{Crouzet} N.,  et~al., 2017, \mn@doi [\aj] {10.3847/1538-3881/153/3/94}, \href
  {http://adsabs.harvard.edu/abs/2017AJ....153...94C} {153, 94}

\bibitem[\protect\citeauthoryear{{Dai} \& {Winn}}{{Dai} \&
  {Winn}}{2017}]{2017AJ....153..205D}
{Dai} F.,  {Winn} J.~N.,  2017, \mn@doi [\aj] {10.3847/1538-3881/aa65d1}, \href
  {http://adsabs.harvard.edu/abs/2017AJ....153..205D} {153, 205}

\bibitem[\protect\citeauthoryear{{Doyle} et~al.,}{{Doyle}
  et~al.}{2013}]{2013MNRAS.428.3164D}
{Doyle} A.~P.,  et~al., 2013, \mn@doi [\mnras] {10.1093/mnras/sts267}, \href
  {http://adsabs.harvard.edu/abs/2013MNRAS.428.3164D} {428, 3164}

\bibitem[\protect\citeauthoryear{{Doyle}, {Davies}, {Smalley}, {Chaplin}  \&
  {Elsworth}}{{Doyle} et~al.}{2014}]{2014MNRAS.444.3592D}
{Doyle} A.~P.,  {Davies} G.~R.,  {Smalley} B.,  {Chaplin} W.~J.,   {Elsworth}
  Y.,  2014, \mn@doi [\mnras] {10.1093/mnras/stu1692}, \href
  {http://adsabs.harvard.edu/abs/2014MNRAS.444.3592D} {444, 3592}

\bibitem[\protect\citeauthoryear{{Enoch}, {Collier Cameron}, {Parley}  \&
  {Hebb}}{{Enoch} et~al.}{2010}]{2010A&A...516A..33E}
{Enoch} B.,  {Collier Cameron} A.,  {Parley} N.~R.,   {Hebb} L.,  2010, \mn@doi
  [\aap] {10.1051/0004-6361/201014326}, \href
  {http://adsabs.harvard.edu/abs/2010A%26A...516A..33E} {516, A33}

\bibitem[\protect\citeauthoryear{{Evans} et~al.,}{{Evans}
  et~al.}{2017}]{2017Natur.548...58E}
{Evans} T.~M.,  et~al., 2017, \mn@doi [\nat] {10.1038/nature23266}, \href
  {http://adsabs.harvard.edu/abs/2017Natur.548...58E} {548, 58}

\bibitem[\protect\citeauthoryear{{Fortney}, {Lodders}, {Marley}  \&
  {Freedman}}{{Fortney} et~al.}{2008}]{2008ApJ...678.1419F}
{Fortney} J.~J.,  {Lodders} K.,  {Marley} M.~S.,   {Freedman} R.~S.,  2008,
  \mn@doi [\apj] {10.1086/528370}, \href
  {http://adsabs.harvard.edu/abs/2008ApJ...678.1419F} {678, 1419}

\bibitem[\protect\citeauthoryear{{Gaia Collaboration} et~al.,}{{Gaia
  Collaboration} et~al.}{2016}]{2016A&A...595A...1G}
{Gaia Collaboration} et~al., 2016, \mn@doi [\aap]
  {10.1051/0004-6361/201629272}, \href
  {http://adsabs.harvard.edu/abs/2016A%26A...595A...1G} {595, A1}

\bibitem[\protect\citeauthoryear{{Gaia Collaboration}, {Brown}, {Vallenari},
  {Prusti}, {de Bruijne}, {Babusiaux}  \& {Bailer-Jones}}{{Gaia Collaboration}
  et~al.}{2018}]{2018arXiv180409365G}
{Gaia Collaboration} {Brown} A.~G.~A.,  {Vallenari} A.,  {Prusti} T.,  {de
  Bruijne} J.~H.~J.,  {Babusiaux} C.,   {Bailer-Jones} C.~A.~L.,  2018,
  preprint, \href {http://adsabs.harvard.edu/abs/2018arXiv180409365G} {}
  (\mn@eprint {arXiv} {1804.09365})

\bibitem[\protect\citeauthoryear{{Gaudi} et~al.,}{{Gaudi}
  et~al.}{2017}]{2017Natur.546..514G}
{Gaudi} B.~S.,  et~al., 2017, \mn@doi [\nat] {10.1038/nature22392}, \href
  {http://adsabs.harvard.edu/abs/2017Natur.546..514G} {546, 514}

\bibitem[\protect\citeauthoryear{{Gibson}, {Nikolov}, {Sing}, {Barstow},
  {Evans}, {Kataria}  \& {Wilson}}{{Gibson} et~al.}{2017}]{2017MNRAS.467.4591G}
{Gibson} N.~P.,  {Nikolov} N.,  {Sing} D.~K.,  {Barstow} J.~K.,  {Evans} T.~M.,
   {Kataria} T.,   {Wilson} P.~A.,  2017, \mn@doi [\mnras]
  {10.1093/mnras/stx353}, \href
  {http://adsabs.harvard.edu/abs/2017MNRAS.467.4591G} {467, 4591}

\bibitem[\protect\citeauthoryear{{Gillon} et~al.,}{{Gillon}
  et~al.}{2010}]{2010A&A...511A...3G}
{Gillon} M.,  et~al., 2010, \mn@doi [\aap] {10.1051/0004-6361/200913507}, \href
  {http://adsabs.harvard.edu/abs/2010A%26A...511A...3G} {511, A3}

\bibitem[\protect\citeauthoryear{{Gillon} et~al.,}{{Gillon}
  et~al.}{2012}]{2012A&A...542A...4G}
{Gillon} M.,  et~al., 2012, \mn@doi [\aap] {10.1051/0004-6361/201218817}, \href
  {http://adsabs.harvard.edu/abs/2012A%26A...542A...4G} {542, A4}

\bibitem[\protect\citeauthoryear{{Gray} \& {Corbally}}{{Gray} \&
  {Corbally}}{2014}]{2014AJ....147...80G}
{Gray} R.~O.,  {Corbally} C.~J.,  2014, \mn@doi [\aj]
  {10.1088/0004-6256/147/4/80}, \href
  {http://adsabs.harvard.edu/abs/2014AJ....147...80G} {147, 80}

\bibitem[\protect\citeauthoryear{{Hartman} et~al.,}{{Hartman}
  et~al.}{2015}]{2015AJ....150..197H}
{Hartman} J.~D.,  et~al., 2015, \mn@doi [\aj] {10.1088/0004-6256/150/6/197},
  \href {http://adsabs.harvard.edu/abs/2015AJ....150..197H} {150, 197}

\bibitem[\protect\citeauthoryear{{Hartman} et~al.,}{{Hartman}
  et~al.}{2016}]{2016AJ....152..182H}
{Hartman} J.~D.,  et~al., 2016, \mn@doi [\aj] {10.3847/0004-6256/152/6/182},
  \href {http://adsabs.harvard.edu/abs/2016AJ....152..182H} {152, 182}

\bibitem[\protect\citeauthoryear{{Hellier} et~al.,}{{Hellier}
  et~al.}{2011}]{2011EPJWC..1101004H}
{Hellier} C.,  et~al., 2011, in European Physical Journal Web of Conferences.
  p. 01004 (\mn@eprint {arXiv} {1012.2286}),
  \mn@doi{10.1051/epjconf/20101101004}

\bibitem[\protect\citeauthoryear{{Heng} \& {Showman}}{{Heng} \&
  {Showman}}{2015}]{2015AREPS..43..509H}
{Heng} K.,  {Showman} A.~P.,  2015, \mn@doi [Annual Review of Earth and
  Planetary Sciences] {10.1146/annurev-earth-060614-105146}, \href
  {http://adsabs.harvard.edu/abs/2015AREPS..43..509H} {43, 509}

\bibitem[\protect\citeauthoryear{{Hirano}, {Suto}, {Winn}, {Taruya}, {Narita},
  {Albrecht}  \& {Sato}}{{Hirano} et~al.}{2011}]{2011ApJ...742...69H}
{Hirano} T.,  {Suto} Y.,  {Winn} J.~N.,  {Taruya} A.,  {Narita} N.,  {Albrecht}
  S.,   {Sato} B.,  2011, \mn@doi [\apj] {10.1088/0004-637X/742/2/69}, \href
  {http://adsabs.harvard.edu/abs/2011ApJ...742...69H} {742, 69}

\bibitem[\protect\citeauthoryear{{Huang} et~al.,}{{Huang}
  et~al.}{2015}]{2015AJ....150...85H}
{Huang} C.~X.,  et~al., 2015, \mn@doi [\aj] {10.1088/0004-6256/150/3/85}, \href
  {http://adsabs.harvard.edu/abs/2015AJ....150...85H} {150, 85}

\bibitem[\protect\citeauthoryear{{Jehin} et~al.,}{{Jehin}
  et~al.}{2011}]{2011Msngr.145....2J}
{Jehin} E.,  et~al., 2011, The Messenger, \href
  {http://adsabs.harvard.edu/abs/2011Msngr.145....2J} {145, 2}

\bibitem[\protect\citeauthoryear{{Johnson} et~al.,}{{Johnson}
  et~al.}{2018}]{2018AJ....155..100J}
{Johnson} M.~C.,  et~al., 2018, \mn@doi [\aj] {10.3847/1538-3881/aaa5af}, \href
  {http://adsabs.harvard.edu/abs/2018AJ....155..100J} {155, 100}

\bibitem[\protect\citeauthoryear{{Kraft}}{{Kraft}}{1967}]{1967ApJ...150..551K}
{Kraft} R.~P.,  1967, \mn@doi [\apj] {10.1086/149359}, \href
  {http://adsabs.harvard.edu/abs/1967ApJ...150..551K} {150, 551}

\bibitem[\protect\citeauthoryear{{Kreidberg} et~al.,}{{Kreidberg}
  et~al.}{2015}]{2015ApJ...814...66K}
{Kreidberg} L.,  et~al., 2015, \mn@doi [\apj] {10.1088/0004-637X/814/1/66},
  \href {http://adsabs.harvard.edu/abs/2015ApJ...814...66K} {814, 66}

\bibitem[\protect\citeauthoryear{{Kreidberg} et~al.,}{{Kreidberg}
  et~al.}{2018a}]{2018AJ....156...17K}
{Kreidberg} L.,  et~al., 2018a, \mn@doi [\aj] {10.3847/1538-3881/aac3df}, \href
  {http://adsabs.harvard.edu/abs/2018AJ....156...17K} {156, 17}

\bibitem[\protect\citeauthoryear{{Kreidberg}, {Line}, {Thorngren}, {Morley}  \&
  {Stevenson}}{{Kreidberg} et~al.}{2018b}]{2018ApJ...858L...6K}
{Kreidberg} L.,  {Line} M.~R.,  {Thorngren} D.,  {Morley} C.~V.,   {Stevenson}
  K.~B.,  2018b, \mn@doi [\apjl] {10.3847/2041-8213/aabfce}, \href
  {http://adsabs.harvard.edu/abs/2018ApJ...858L...6K} {858, L6}

\bibitem[\protect\citeauthoryear{{Lund} et~al.,}{{Lund}
  et~al.}{2017}]{2017AJ....154..194L}
{Lund} M.~B.,  et~al., 2017, \mn@doi [\aj] {10.3847/1538-3881/aa8f95}, \href
  {http://adsabs.harvard.edu/abs/2017AJ....154..194L} {154, 194}

\bibitem[\protect\citeauthoryear{{Maxted} et~al.,}{{Maxted}
  et~al.}{2011}]{2011PASP..123..547M}
{Maxted} P.~F.~L.,  et~al., 2011, \mn@doi [\pasp] {10.1086/660007}, \href
  {http://adsabs.harvard.edu/abs/2011PASP..123..547M} {123, 547}

\bibitem[\protect\citeauthoryear{{Maxted}, {Serenelli}  \&
  {Southworth}}{{Maxted} et~al.}{2015}]{2015A&A...575A..36M}
{Maxted} P.~F.~L.,  {Serenelli} A.~M.,   {Southworth} J.,  2015, \mn@doi [\aap]
  {10.1051/0004-6361/201425331}, \href
  {http://adsabs.harvard.edu/abs/2015A%26A...575A..36M} {575, A36}

\bibitem[\protect\citeauthoryear{{Mazeh}, {Perets}, {McQuillan}  \&
  {Goldstein}}{{Mazeh} et~al.}{2015}]{2015ApJ...801....3M}
{Mazeh} T.,  {Perets} H.~B.,  {McQuillan} A.,   {Goldstein} E.~S.,  2015,
  \mn@doi [\apj] {10.1088/0004-637X/801/1/3}, \href
  {http://adsabs.harvard.edu/abs/2015ApJ...801....3M} {801, 3}

\bibitem[\protect\citeauthoryear{{Munari} \& {Zwitter}}{{Munari} \&
  {Zwitter}}{1997}]{1997A&A...318..269M}
{Munari} U.,  {Zwitter} T.,  1997, \aap, \href
  {http://adsabs.harvard.edu/abs/1997A%26A...318..269M} {318, 269}

\bibitem[\protect\citeauthoryear{{Parmentier} et~al.,}{{Parmentier}
  et~al.}{2018}]{2018arXiv180500096P}
{Parmentier} V.,  et~al., 2018, preprint, \href
  {http://adsabs.harvard.edu/abs/2018arXiv180500096P} {} (\mn@eprint {arXiv}
  {1805.00096})

\bibitem[\protect\citeauthoryear{{Pepe} et~al.,}{{Pepe}
  et~al.}{2002}]{2002Msngr.110....9P}
{Pepe} F.,  et~al., 2002, The Messenger, \href
  {http://adsabs.harvard.edu/abs/2002Msngr.110....9P} {110, 9}

\bibitem[\protect\citeauthoryear{{Pollacco} et~al.,}{{Pollacco}
  et~al.}{2008}]{2008MNRAS.385.1576P}
{Pollacco} D.,  et~al., 2008, \mn@doi [\mnras]
  {10.1111/j.1365-2966.2008.12939.x}, \href
  {http://adsabs.harvard.edu/abs/2008MNRAS.385.1576P} {385, 1576}

\bibitem[\protect\citeauthoryear{{Pont}, {Husnoo}, {Mazeh}  \&
  {Fabrycky}}{{Pont} et~al.}{2011}]{2011MNRAS.414.1278P}
{Pont} F.,  {Husnoo} N.,  {Mazeh} T.,   {Fabrycky} D.,  2011, \mn@doi [\mnras]
  {10.1111/j.1365-2966.2011.18462.x}, \href
  {http://adsabs.harvard.edu/abs/2011MNRAS.414.1278P} {414, 1278}

\bibitem[\protect\citeauthoryear{{Queloz} et~al.,}{{Queloz}
  et~al.}{2001}]{2001Msngr.105....1Q}
{Queloz} D.,  et~al., 2001, The Messenger, \href
  {http://cdsads.u-strasbg.fr/abs/2001Msngr.105....1Q} {105, 1}

\bibitem[\protect\citeauthoryear{{Sestito} \& {Randich}}{{Sestito} \&
  {Randich}}{2005}]{2005A&A...442..615S}
{Sestito} P.,  {Randich} S.,  2005, \mn@doi [\aap]
  {10.1051/0004-6361:20053482}, \href
  {http://adsabs.harvard.edu/abs/2005A%26A...442..615S} {442, 615}

\bibitem[\protect\citeauthoryear{{Siverd} et~al.,}{{Siverd}
  et~al.}{2018}]{2018AJ....155...35S}
{Siverd} R.~J.,  et~al., 2018, \mn@doi [\aj] {10.3847/1538-3881/aa9e4d}, \href
  {http://adsabs.harvard.edu/abs/2018AJ....155...35S} {155, 35}

\bibitem[\protect\citeauthoryear{{Smalley} et~al.,}{{Smalley}
  et~al.}{2011}]{2011A&A...526A.130S}
{Smalley} B.,  et~al., 2011, \mn@doi [\aap] {10.1051/0004-6361/201015992},
  \href {http://adsabs.harvard.edu/abs/2011A%26A...526A.130S} {526, A130}

\bibitem[\protect\citeauthoryear{{Stassun} \& {Torres}}{{Stassun} \&
  {Torres}}{2018}]{2018ApJ...862...61S}
{Stassun} K.~G.,  {Torres} G.,  2018, \mn@doi [\apj]
  {10.3847/1538-4357/aacafc}, \href
  {http://adsabs.harvard.edu/abs/2018ApJ...862...61S} {862, 61}

\bibitem[\protect\citeauthoryear{{Stevenson}, {Bean}, {Madhusudhan}  \&
  {Harrington}}{{Stevenson} et~al.}{2014}]{2014ApJ...791...36S}
{Stevenson} K.~B.,  {Bean} J.~L.,  {Madhusudhan} N.,   {Harrington} J.,  2014,
  \mn@doi [\apj] {10.1088/0004-637X/791/1/36}, \href
  {http://adsabs.harvard.edu/abs/2014ApJ...791...36S} {791, 36}

\bibitem[\protect\citeauthoryear{{Talens} et~al.,}{{Talens}
  et~al.}{2017}]{2017A&A...606A..73T}
{Talens} G.~J.~J.,  et~al., 2017, \mn@doi [\aap] {10.1051/0004-6361/201731282},
  \href {http://adsabs.harvard.edu/abs/2017A%26A...606A..73T} {606, A73}

\bibitem[\protect\citeauthoryear{{Talens} et~al.,}{{Talens}
  et~al.}{2018}]{2018A&A...612A..57T}
{Talens} G.~J.~J.,  et~al., 2018, \mn@doi [\aap] {10.1051/0004-6361/201731512},
  \href {http://adsabs.harvard.edu/abs/2018A%26A...612A..57T} {612, A57}

\bibitem[\protect\citeauthoryear{{Temple} et~al.,}{{Temple}
  et~al.}{2017}]{2017MNRAS.471.2743T}
{Temple} L.~Y.,  et~al., 2017, \mn@doi [\mnras] {10.1093/mnras/stx1729}, \href
  {http://adsabs.harvard.edu/abs/2017MNRAS.471.2743T} {471, 2743}

\bibitem[\protect\citeauthoryear{{Torres}, {Andersen}  \&
  {Gim{\'e}nez}}{{Torres} et~al.}{2010}]{2010A&ARv..18...67T}
{Torres} G.,  {Andersen} J.,   {Gim{\'e}nez} A.,  2010, \mn@doi [\aapr]
  {10.1007/s00159-009-0025-1}, \href
  {http://adsabs.harvard.edu/abs/2010A%26ARv..18...67T} {18, 67}

\bibitem[\protect\citeauthoryear{{Triaud}}{{Triaud}}{2017}]{2017haex.bookE...2T}
{Triaud} A.~H.~M.~J.,  2017, {The Rossiter-McLaughlin Effect in Exoplanet
  Research}.
p.~2, \mn@doi{10.1007/978-3-319-30648-3_2-1}

\bibitem[\protect\citeauthoryear{{Valsecchi} \& {Rasio}}{{Valsecchi} \&
  {Rasio}}{2014}]{2014ApJ...786..102V}
{Valsecchi} F.,  {Rasio} F.~A.,  2014, \mn@doi [\apj]
  {10.1088/0004-637X/786/2/102}, \href
  {http://adsabs.harvard.edu/abs/2014ApJ...786..102V} {786, 102}

\bibitem[\protect\citeauthoryear{{Weiss} \& {Schlattl}}{{Weiss} \&
  {Schlattl}}{2008}]{2008Ap&SS.316...99W}
{Weiss} A.,  {Schlattl} H.,  2008, \mn@doi [\apss] {10.1007/s10509-007-9606-5},
  \href {http://adsabs.harvard.edu/abs/2008Ap%26SS.316...99W} {316, 99}

\bibitem[\protect\citeauthoryear{{Winn}, {Fabrycky}, {Albrecht}  \&
  {Johnson}}{{Winn} et~al.}{2010}]{2010ApJ...718L.145W}
{Winn} J.~N.,  {Fabrycky} D.,  {Albrecht} S.,   {Johnson} J.~A.,  2010, \mn@doi
  [\apjl] {10.1088/2041-8205/718/2/L145}, \href
  {http://adsabs.harvard.edu/abs/2010ApJ...718L.145W} {718, L145}

\bibitem[\protect\citeauthoryear{{Wyttenbach}, {Ehrenreich}, {Lovis}, {Udry}
  \& {Pepe}}{{Wyttenbach} et~al.}{2015}]{2015A&A...577A..62W}
{Wyttenbach} A.,  {Ehrenreich} D.,  {Lovis} C.,  {Udry} S.,   {Pepe} F.,  2015,
  \mn@doi [\aap] {10.1051/0004-6361/201525729}, \href
  {http://adsabs.harvard.edu/abs/2015A%26A...577A..62W} {577, A62}

\bibitem[\protect\citeauthoryear{{Yan} \& {Henning}}{{Yan} \&
  {Henning}}{2018}]{2018NatAs.tmp...86Y}
{Yan} F.,  {Henning} T.,  2018, \mn@doi [Nature Astronomy]
  {10.1038/s41550-018-0503-3}, \href
  {http://adsabs.harvard.edu/abs/2018NatAs.tmp...86Y} {}

\bibitem[\protect\citeauthoryear{{Zhou} et~al.,}{{Zhou}
  et~al.}{2016a}]{2016AJ....152..136Z}
{Zhou} G.,  et~al., 2016a, \mn@doi [\aj] {10.3847/0004-6256/152/5/136}, \href
  {http://adsabs.harvard.edu/abs/2016AJ....152..136Z} {152, 136}

\bibitem[\protect\citeauthoryear{{Zhou}, {Latham}, {Bieryla}, {Beatty},
  {Buchhave}, {Esquerdo}, {Berlind}  \& {Calkins}}{{Zhou}
  et~al.}{2016b}]{2016MNRAS.460.3376Z}
{Zhou} G.,  {Latham} D.~W.,  {Bieryla} A.,  {Beatty} T.~G.,  {Buchhave} L.~A.,
  {Esquerdo} G.~A.,  {Berlind} P.,   {Calkins} M.~L.,  2016b, \mn@doi [\mnras]
  {10.1093/mnras/stw1107}, \href
  {http://adsabs.harvard.edu/abs/2016MNRAS.460.3376Z} {460, 3376}

\bibitem[\protect\citeauthoryear{{Zhou} et~al.,}{{Zhou}
  et~al.}{2017}]{2017AJ....153..211Z}
{Zhou} G.,  et~al., 2017, \mn@doi [\aj] {10.3847/1538-3881/aa674a}, \href
  {http://adsabs.harvard.edu/abs/2017AJ....153..211Z} {153, 211}

\makeatother
\end{thebibliography}


\bsp
\label{lastpage}
\end{document}